\documentclass[12pt]{article}
\usepackage{amsmath}
\usepackage{amssymb}
\usepackage{amsthm}
\usepackage{fullpage}
\usepackage[utf8]{inputenc}
\usepackage{mathtools}
\usepackage{url}
\usepackage[usenames,svgnames]{xcolor}

\usepackage[pagebackref]{hyperref} 
\usepackage[all]{hypcap} 

\theoremstyle{plain}

\theoremstyle{definition}

\numberwithin{equation}{section} 

\hypersetup{
  breaklinks,colorlinks,
  citecolor=Green,
  linkcolor=BlueViolet,
  urlcolor=DarkBlue,
  pdftitle={2D Toda tau-functions as combinatorial generating functions},
  pdfauthor={Mathieu Guay-Paquet and John Harnad},
}

\begin{document}
\baselineskip 16pt

\medskip
\begin{center}
\begin{Large}\fontfamily{cmss}
\fontsize{17pt}{27pt}
\selectfont
\textrm{Comment on ``Higher order Painlev\'e equations and their symmetries via reductions of a class of integrable models''}
\end{Large}\\
\bigskip
\begin{large} {Svinin Andrei K.}
 \end{large}
\\
\bigskip
\begin{small}
{\em Institute for System Dynamics and Control Theory\\
Siberian Branch of Russian Academy of Sciences \\}
svinin@icc.ru \\
\end{small}
\end{center}
\bigskip

This comment is written  to compare the results of our two papers \cite{Svinin1} and \cite{Svinin2} with the results of \cite{Aratyn}. 
In \cite{Svinin1} we have been proposed  a modified version of Krichever-Dickey rational reductions of the KP hierarchy \cite{Dickey}, \cite{Krichever} with corresponding lattice representation. To make our comment more clear, let us first give a short description of this approach. 

Given any positive integers $n\geq 1$ and $p>n$, corresponding integrable hierarchy  is defined by two evolution equations 
\begin{equation}
\partial_sG_{(k)}=\left(\mathcal{L}_{(k+1)}^{s}\right)_{+}G_{(k)}-G_{(k)}\left(\mathcal{L}_{(k)}^{s}\right)_{+},
\label{ev1}
\end{equation}
\begin{equation}
\partial_sH_{(k)}=\left(\mathcal{M}_{(k+1)}^{s}\right)_{+}H_{(k)}-H_{(k)}\left(\mathcal{M}_{(k)}^{s}\right)_{+}
\label{ev2}
\end{equation}
with
\[
G_{(k)}\equiv\partial + v_k,\;\;
H_{(k)}\equiv\partial + u_k\;\; 
\]
and finite collection of pseudo-differential  operators $\left\{\mathcal{L}_{(j)}, \mathcal{M}_{(j)}\right\}$ defined through
\[
\mathcal{L}_{(j)}^{p-n}=G_{(j-1)}\cdots G_{(1)}H_{(1)}^{-1}\cdots H_{(n)}^{-1}G_{(p)}\cdots G_{(j)},\;\;
j=1,\ldots, p+1,
\]
\[
\mathcal{M}_{(j)}^{p-n}=H_{(j)}^{-1}\cdots H_{(n)}^{-1}G_{(p)}\cdots G_{(1)}H_{(1)}^{-1}\cdots H_{(j-1)}^{-1},\;\;
j=1,\ldots, n+1.
\]
As a consequence of (\ref{ev1}) and (\ref{ev2}), pseudo-differential operators $L=P_n^{-1}Q_p$ and $M=Q_pP_n^{-1}$ with factorized operators
\[
Q_p=G_{(p)}\cdots G_{(1)},\;\;
P_n=H_{(n)}\cdots H_{(1)} 
\]
satisfy KP Lax equations. The  fields $v_j$ and $u_j$, by virtue of their definition, must satisfy the condition
\begin{equation}
\sum_{j=1}^pv_j=\sum_{j=1}^nu_j.
\label{condition}
\end{equation}
This means that in fact one has $n+p-1$ independent fields.
Let us also write down evolution equations governing  the second flow in the hierarchy in explicit form. We get from (\ref{ev1}) and (\ref{ev2}) that
\begin{equation}
\partial_2v_k=\left(v_k^{\prime}-v_k^2\right)^{\prime}+\frac{2}{p-n}\left(\sum_{j=1}^{n-1}(n-j)u_j^{\prime}-\sum_{j=1}^{k-1}(n-j)v_j^{\prime}-\sum_{j=k}^{p-1}(p-j)v_j^{\prime}\right)^{\prime}-R^{\prime}, 
\label{1}
\end{equation}
\begin{equation}
\partial_2u_k=\left(u_k^{\prime}-u_k^2\right)^{\prime}+\frac{2}{p-n}\left(-\sum_{j=1}^{p-1}(p-j)v_j^{\prime}+\sum_{j=1}^{k-1}(p-j)u_j^{\prime}+\sum_{j=k}^{n-1}(n-j)u_j^{\prime}\right)^{\prime}-R^{\prime} 
\label{2}
\end{equation}
with
\[
R\equiv \frac{2}{p-n}\left(\sum_{j<k}v_jv_k-\sum_{j<k}u_ju_k\right).
\]

In the paper \cite{Aratyn}, the authors consider the particular case $p=n+1$ with $n=M$ of modified rational reductions of KP hierarchy given by (\ref{ev1}) and (\ref{ev2}). Remark that the second flow in this case is yielding by the system of evolution equations
\[
\partial_2v_k=\left(v_k^{\prime}-v_k^2\right)^{\prime}+2\left(\sum_{j=1}^{n-1}(n-j)u_j^{\prime}-\sum_{j=1}^{k-1}(n-j)v_j^{\prime}-\sum_{j=k}^{n}(n-j+1)v_j^{\prime}\right)^{\prime}-R^{\prime}, 
\]
\[
\partial_2u_k=\left(u_k^{\prime}-u_k^2\right)^{\prime}+2\left(-\sum_{j=1}^{n}(n-j+1)v_j^{\prime}+\sum_{j=1}^{k-1}(n-j+1)u_j^{\prime}+\sum_{j=k}^{n-1}(n-j)u_j^{\prime}\right)^{\prime}-R^{\prime} 
\]
which is a specification of (\ref{1}) and (\ref{2}). Clearly, in this case one has $2M$ independent fields.   Lax operator (2.10) denoted by $L_M$ in \cite{Aratyn} is of the form $Q_{M+1}P_M^{-1}$ with factorized operators $P_M$ and $Q_{M+1}$ and it involves a finite number of the fields $\{c_j, e_j : k=j,\ldots, M\}$ in such a way that our fields $v_k$ and $u_k$ turn out linearly depend on $\{c_j, e_j\}$ and moreover the condition (\ref{condition}) becomes an identity and can be discarded. For example, in the case $M=2$, it can be computed to obtain
\[
v_1=-c_1-c_2,\;\;
v_1=-e_1-c_2,\;\;
v_1=-e_2,\;\;
\]
\[
u_1=-e_1-c_1-c_2,\;\;
v_1=-e_2-c_2.\;\;
\]

The authors of \cite{Aratyn} write down the second flow in explicit form (3.1) making use the corresponding Hamiltonian coming from Lax operator. 
One has only to note that their equations differ from our equations in sign of right-hand sides.
As for discrete symmetry, transformation $g$ in \cite{Aratyn}  is in fact discrete symmetry transformation $s_1$ in \cite{Svinin1} which in general case is generated by differential-difference equation
\begin{equation}
\sum_{j=1}^{p-n}a_{i+j-1}^{\prime}=\sum_{j=1}^{p-n}a_{i+j-1}\left(\sum_{j=1}^{n}a_{i+j-n-1}-\sum_{j=1}^{n}a_{i+j+p-n-1}\right)
\label{dde}
\end{equation}
and responsible for  the shift $i\rightarrow i+n$. In the particular case $p=n+1$, in an obvious way, equation (\ref{dde}) becomes well-known Itoh-Narita-Bogoyavlenskii (INB) lattice \cite{Itoh}, \cite{Narita}, \cite{Bogoyavlenskii}
\begin{equation}
a_i^{\prime}=a_i\left(\sum_{j=1}^na_{i-j}-\sum_{j=1}^na_{i+j}\right)
\label{INB}
\end{equation}
which governs discrete symmetries of corresponding integrable hierarchies of evolution differential equations.

In section 6 of \cite{Svinin2}, we have been shown that self-similarity reduction of INB lattice hierarchy leads to Painlev\'e equations of type $A^{(1)}_{2n}$. Simplifying the situation, we may consider INB lattice (\ref{INB}) supplemented by the condition
\begin{equation}
a_i\left(\sum_{j=-n}^na_{i+j}-x\right)=\alpha_i,\;\;
\sum_{j=1}^n\alpha_{i-j}=\sum_{j=1}^n\alpha_{i+j}+1
\label{constraint}
\end{equation}
which follows in fact from self-similarity constraint $\left(1+x\partial+2t_2\partial_2\right)a_i=0.$ As was shown in \cite{Svinin1},  that stationary version of (\ref{constraint}), for all $n\geq 1$, is an integrable discretization of the $P_I$ equation $w^{\prime\prime}=6w^2+t$. In subsection 4.2 of the work \cite{Svinin2}, it was shown that INB lattice (\ref{INB}) together with compatible constraint (\ref{constraint}) is equivalent to Painlev\'e equations of type $A^{(1)}_{2n}$ \cite{Noumi}. It is natural that the authors of \cite{Aratyn} using the self-similarity reduction of their evolution equations also come to $A^{(1)}_{2n}$ Painlev\'e equations.


\end{document}